\documentclass[journal=jpclett,manuscript=communication]{achemso}
\usepackage[version=3]{mhchem} 
\usepackage[T1]{fontenc}       
\usepackage{longtable}
\usepackage{multirow}
\usepackage{array}
\usepackage{booktabs} 
\usepackage{amsmath}
\usepackage{graphicx}
\usepackage{subfig}

\author{Srimanta Pakhira}
\email{spakhira@iiti.ac.in}
\affiliation{Discipline of Physics,  Indian Institute of Technology Indore, Simrol Campus, Khandwa Road, Simrol,  Indore-453552, Madhya Pradesh, India.}
\alsoaffiliation{Discipline of Metallurgy Engineering and Materials Science, Indian Institute of Technology Indore, Simrol Campus, Khandwa Road, Simrol,  Indore-453552, Madhya Pradesh, India.}

\author{Jose L. Mendoza-Cortes} 
\email{mendoza@eng.famu.fsu.edu}
\phone{+1-850-410-6298}
\fax{+1-850-410-6150}
\affiliation{Department of Chemical \& Biomedical Engineering, Florida A\&M University - Florida State University (FAMU-FSU), Joint College of Engineering, Tallahassee, FL, 32310, USA.}
\alsoaffiliation{Condensed Matter Theory, National High Magnetic Field Laboratory (NHMFL), Florida State University, Tallahassee, FL, 32310, USA.}
\alsoaffiliation{Department of Scientific Computing, Materials Science and Engineering, High Performance Materials Institute, Florida State University, Tallahassee, FL, 32310, USA.}
\alsoaffiliation{Department of Physics, Florida State University, Tallahassee, FL, 32306, USA.}  

\date{\today} 
\title {The Quantum Nature in the Interaction of Molecular Hydrogen with Porous Materials: Implications for Practical Hydrogen Storage} 

\begin{document}

\begin{abstract}
	
The storage of hydrogen (\ce{H2}) is of economic and ecological relevance, because it could potentially replace petroleum-based fuels. However, \ce{H2} storage at mild condition remains one of the bottlenecks for its widespread usage. In order to devise successful \ce{H2} storage strategies, there is a need for a fundamental understanding of the weak and elusive hydrogen interactions  at the quantum mechanical level. One of the most promising strategies for storage at mild pressure and temperature is physisorption. Porous materials are specially effective at physisorption, however the process at the quantum level has been under-studied.  Here, we present quantum calculations to study the interaction of \ce{H2} with building units of porous materials. We report 240 \ce{H2} complexes made of different transition metal (Tm) atoms, chelating ligands, spins, oxidation states, and geometrical configurations. We found that both the dispersion and electrostatics interactions are the major contributors to the interaction energy between \ce{H2} and the transition metal complexes. The binding energy for some of these complexes is in the range of at least 10 kJ/mol for many interactions sites, which is one of these main requirements for practical \ce{H2} storage. Thus, these results are of fundamental nature for practical \ce{H2} storage in porous materials.
	  
\end{abstract}

\section{Introduction} 
The world's increasing energy demands, limited petroleum feed stocks and increasing greenhouse gas emissions are forcing us to restructure our energy economy towards sustainable and renewable energy sources. Finding low cost, safe, and efficient energy storage materials is a major milestone toward developing renewable energy technology which can potentially replace the carbon-based fossil fuels. In this context, molecular hydrogen, or \ce{H2} for short, with an energy content of 142 MJ kg$^{-1}$ is an ideal and widely accepted green fuel because of its environmental friendliness, and sustainability. \ce{H2} has an energy density much greater than gasoline and emits no green house gases such as carbon dioxide (\ce{CO2}) or carbon monooxide (\ce{CO}) after burning. One of the biggest challenges to reach practical applications is to achieve high density hydrogen storage at mild conditions. Free hydrogen does not occur naturally in large quantities, and it should be generated from some other renewable energy sources, \textit{e.g.} artificial photosynthesis.\cite{Lucht2015} In other words, \ce{H2} is an energy carrier (like electricity), not a primary energy source (like coal). For the advancement of hydrogen technologies to be used in transportation and other many applications; the research on hydrogen production, storage, and transformation should be further developed. Thus, hydrogen storage is a key enabling technology. Accordingly an energy efficient method for the storage of \ce{H2} is a necessary technology for its effective use as a fuel.

Recently, several studies and investigations showed that the addition of a transition metal (Tm) atom inside porous materials increases the total capacity of \ce{H2} storage.\cite{mendoza_design_2016,yildirim_direct_2005, arter_fivefold_2016} A reversible mechanism for adsorption and release of \ce{H2} at mild conditions is needed for any practical storage application, which can be achieved with physisorption. Some examples of storage materials capable of physisorption are: Metal-Organic Frameworks (MOFs), Covalent Organic Frameworks (COFs), Zeolites, to mention a few. It has been hypothesized that the ideal range for the heat of adsorption ($Q_{st}$) is around 7-15 kJ/mol for efficient charge/discharge physisorption at ambient temperature (233-258 K).\cite{Aduenko2018} Recent work has shown that the $Q_{st}$ can be approximated by the binding enthalpy ($\Delta \textit{H}_{\textrm{bind}}^{\circ}$) computed by first-principles calculations.\cite{mendoza_design_2016} However, the nature of the interactions of the molecular components with \ce{H2} has not been studied in detail. Accordingly, we present a study of the interactions between \ce{H2} with the host. Current materials reach heats of adsorption of less than 8 kJ/mol and decay as the first sorption sites are saturated at ambient temperature.\cite{Han_JACS_2007,Bhatia_Langmuir_2006} The initial studies made use mostly of dispersion interaction which are weak. The orbital interactions are the strongest in these Tm-linker complexes. At the fundamental level, \ce{H2} can interact with other atoms, molecules, and solids via electrostatics, dispersion, and orbital interactions.  \cite{Lochan_PCCP_2006,Pascal_JPCL_2011,Kubas_ChemRev_2007} However, the nature of \ce{H2} interaction with $d-$orbitals of the chelated Tm atoms inside a nanoporous is not well established yet.       

\begin{figure}[htbp!]  
	\includegraphics[width=.99\linewidth]{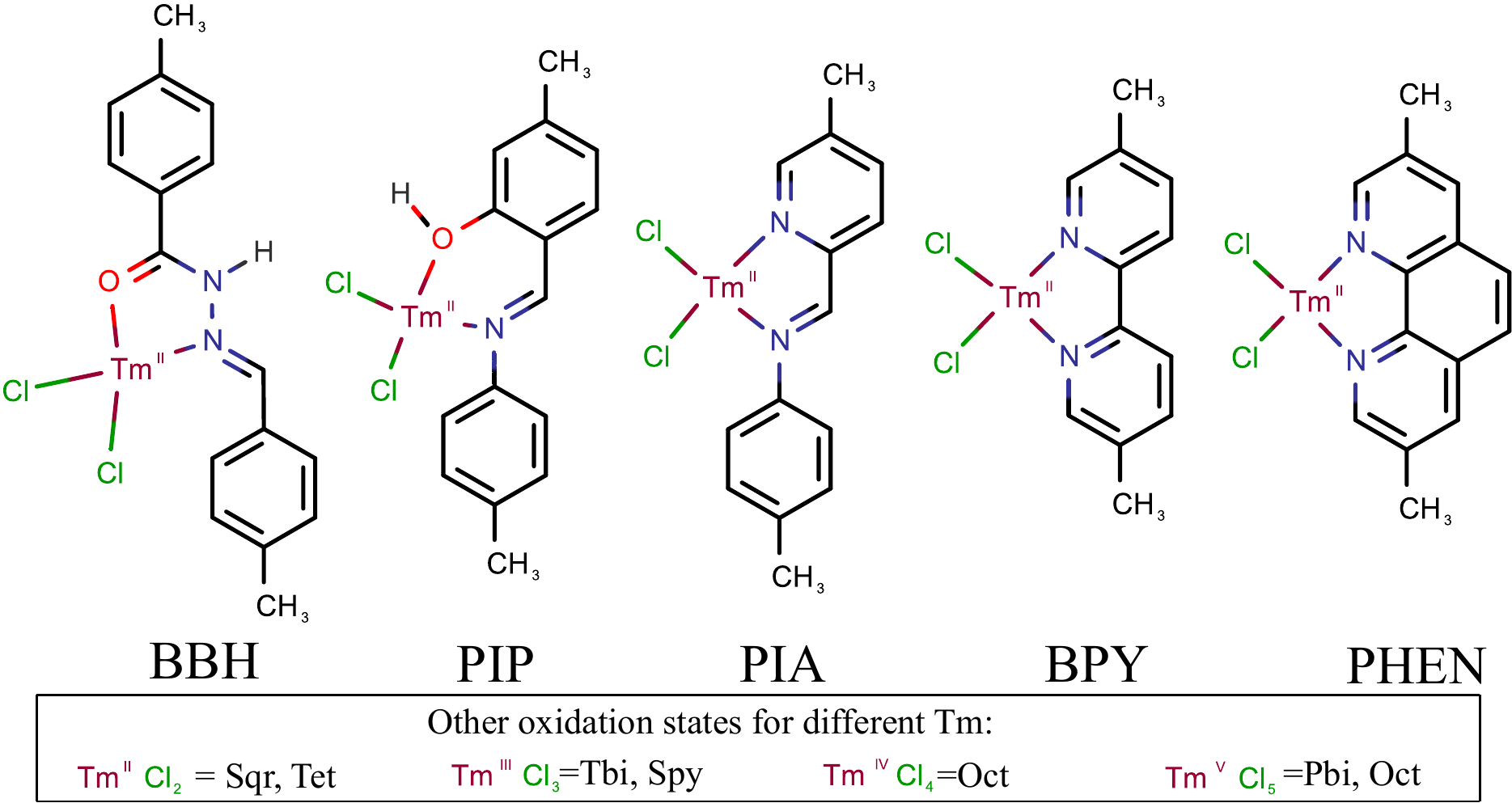}
	\caption{\label{fig:FigureLigands} Ligands used as linkers in the design of crystalline porous materials which contain plausible transition metal binding sites. Notice the change of geometry for different oxidation states.}
\end{figure}

In this communication, the fundamental origin of the \ce{H2} interactions with chelated transition metals is explained based on first-principles calculations.\cite{Becke1988, Lee1988} Thus, we used unrestricted B3LYP-D3 level of theory that takes into account dispersion and orbital interactions of the Tm.\cite{Grimme2006, Grimme2010, Jose2018jpcc, Pakhira2017c, Masataka2015a, Jingshu2018, Niu2018x, Jose2018jcp} The detailed computational methods are described in the Supplementary Information. The primary goal of this paper is to present a fundamental investigation into the interaction between hydrogen and organic linkers to understand the chemical principles which influence the overall adsorption and storage, which can be used in COFs, MOFs or other porous materials. Crystalline porous materials such as MOFs and COFs are linked by organic ligands (also known as `linker'), which can integrate organic units with atomic precision into periodic structures.\cite{Pakhira2017c} The linkers studied are shown in Figure \ref{fig:FigureLigands}: (E)-N'-benzylidene-benzohydrazide (\textbf{BBH}), 
(E)-2- ((phenylimino) methyl) phenol (\textbf{PIP}), 
(E)-N-(pyridin-2-ylmethylene) aniline (\textbf{PIA}), 
2,2'-bipyridine (\textbf{BPY})\cite{Ghosh_InorgChem_2008,Wu_JChemSoc_1999}, and phenanthroline (\textbf{PHEN}). 
These linkers have been chosen because they are reported experimentally and some of these linkers have been used for some COFs and MOFs already. We studied all the expected geometries: square planar (Sqr), tetrahedral (Tet), trigonal bipyramidal (Tbi), square pyramidal (Spy), octahedral (Oct), and pentagonal bipyramidal (Pbi). We studied the most common oxidation state of the Tm atoms which are noted in parenthesis: Sc(III), V(V), Ti(IV), Cr(III), Mn(II), Fe(II), Co(II), Ni(II), Cu(II), and precious transition metals: Pd(II) and Pt(II). Different spins states configurations were calculated for each Tm complex (See SI). If these compounds and interactions are understood well, then the design principles for \ce{H2} storage materials will follow. We expect that this study will be able to provide some guidelines for the preparation of future successful \ce{H2} adsorbing linkers, which will offer more \ce{H2} storage. 

\begin{figure}[htbp!]
	\includegraphics[width=.99\linewidth]{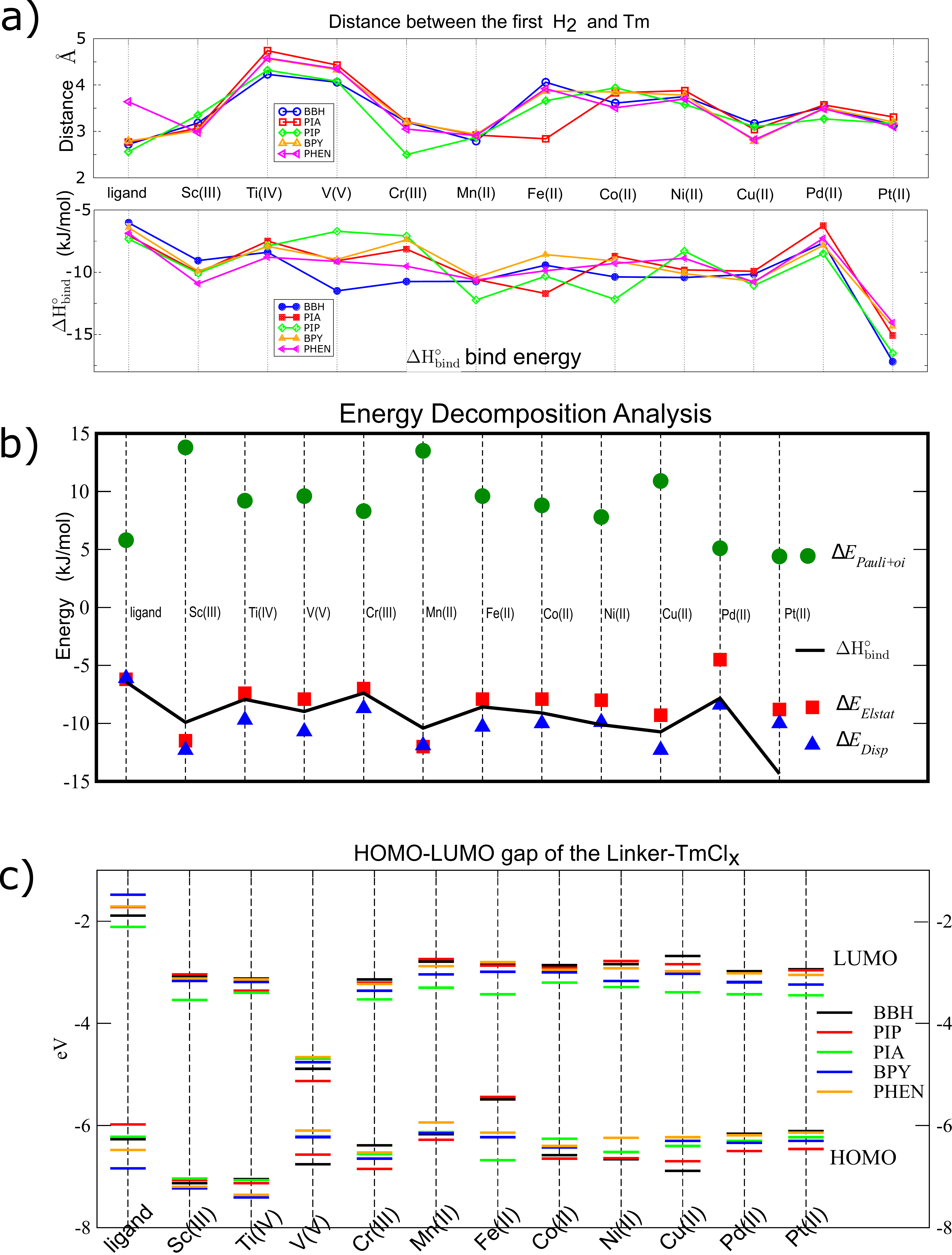}
	\caption{\label{fig:distanceTm_H2} (a) The distance between chelated Tm with linkers (\textbf{BBH, PIA, PIP, BPY, PHEN}) to the first \ce{H2} molecule (top) and the binding energy (bottom); (b) their energy decomposition analysis; and (c) HOMO-LUMO gap.}
\end{figure}

\begin{figure}[htbp!]
    \includegraphics[width=.99\linewidth]{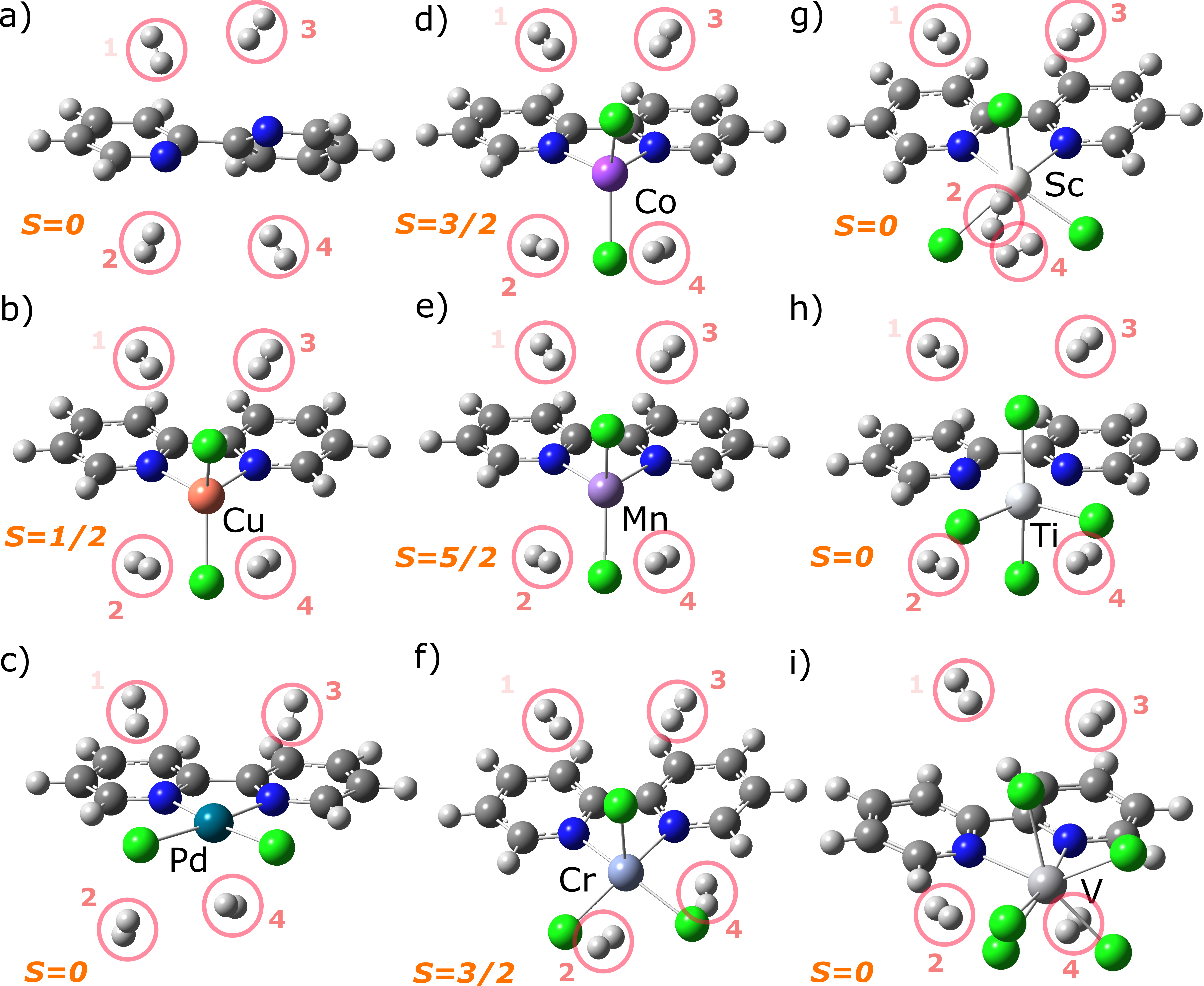}
	\caption{\label{fig:3Dmolecules}The configurations \textbf{BPY}-$\mathrm{TmCl_x}$ complexes interacting with the first 4 \ce{H2} are shown: (a) ligand alone; (b) \textbf{BPY}-\ce{CuCl_2}, Cu(II) with $S$=1/2; (c) \textbf{BPY}-\ce{PdCl_2}, Pd(II) with $S$ = 0; (d) \textbf{BPY}-\ce{CoCl_2}, Co(II) with $S$ = 3/2; (e) \textbf{BPY}-\ce{MnCl_2}, Mn(II) with $S$ = 5/2; (f) \textbf{BPY}-\ce{CrCl_3}, Cr(III) with $S$=3/2 forms a square pyramidal geometry; (g) \textbf{BPY}-\ce{ScCl_3}, Sc (III) with $S$=0 forms a trigonal bipyramidal; (h)\textbf{BPY}-\ce{TiCl_4}, Ti(IV) with $S$ = 0 has octahedral geometry; and (i)\textbf{BPY}-\ce{VCl_5},  V(V) with $S$ = 0 forms a pentagonal bipyramidal geometry.}
\end{figure}

The $1^{st}$ studied parameter is the distance of chelated Tm to the centroid of the first absorbed \ce{H2} molecule as shown in FIG \ref{fig:distanceTm_H2}a (Top).  In pristine ligands, without any Tm, the distances are measured to the centroid of the binding site of the Tm; i.e. between O-N or N-N. Our calculations showed that Cr(III) has the shortest distance between the Tm and the pristine ligands, which suggest that the interaction is among the strongest of all the first row Tm atoms. The distance between the \ce{H2} molecule and any particular Tm  varies only slightly with the type of ligand. However, the magnitude of the binding enthalpy increases as the distance between Tm and \ce{H2} decreases. In the case of \textbf{PIP}  (Figure \ref{fig:distanceTm_H2}a);  the binding sites contain \ce{-OH} where the H atom may rotate. This H atom rotation may explain the higher variation in binding enthalpy  observed in \textbf{PIP} compared to other ligands. We found that the linkers chelated with Pt(II) have the most negative value of \ce{H2} binding enthalpy, $\Delta \textit{H}_{\textrm{bind}}^{\circ}$, which indicates they form the most stable complexes among all chelated compounds studied here. The average value of the binding energy for these Tm complexes is very close to the desired ideal $\Delta \textit{H}_{\textrm{bind}}^{\circ}$ for reversible physisorption of 7-15 kJ/mol. Thus, our present study demonstrates that the linkers chelated with Tm will bind \ce{H2} more strongly. The linkers discussed are based on widely used chelating groups in coordination chemistry and can be used as building blocks for future porous materials.

The $2^{nd}$ studied parameter is the  geometry which depends on the total spin number $S$ and coordination number.\cite{thonhauser_spin_2015,sun_effect_2007} The optimized structures of the chelated, \textbf{BPY}, interacting with four \ce{H2} molecules  are shown in Figure \ref{fig:3Dmolecules}. The numbers next to the \ce{H2} molecules represent the sequence of the \ce{H2} addition to the \textbf{BPY}-\ce{TmCl_x}.  We choose \textbf{BPY} for the configuration analysis because \textbf{BPY} is highly symmetric, therefore the local minima sites for the \ce{H2} are unrelated to other parameters such as dihedral angle of the linker-\ce{TmCl_x} complex.  For the ligand alone, the \ce{H2} molecules  form symmetrical configurations relative to the binding site (Figure \ref{fig:3Dmolecules}a).

In the chelated-Tm with the square geometry: Cu(II), Pd(II), and Pt(II), the first two \ce{H2} molecules are located at the open sites of the Tm, which are above and below the Tm (Figure \ref{fig:3Dmolecules}b), while the third \ce{H2} displaces the first \ce{H2} from its initial location. Stronger binding enthalpy is observed for any Tm, in the tetrahedral geometry than in the square planar geometry. For example, in the \textbf{BBH}-\ce{NiCl2} complex, the Ni(II) has spin \textit{S}=2/2 in the tetrahedral geometry with a binding enthalpy of 1.2 kJ/mol more negative than the square planar geometry with spin \textit{S}=0. On the other hand, \textbf{PIP}-\ce{CoCl2} with spin $S=3/2$ and $S=1/2$ are in a tetrahedral geometry and have \ce{H2} similar binding enthalpies with less than $0.1$ kJ/mol difference. Only the Tm with $S$=0 can have a perfect square geometry; \textit{e.g.} Cu(II) with \textit{S}=1/2 forms a distorted square geometry compound. In Co(II), Fe(II), Mn(II), and Ni(II) with the tetrahedral geometry  the first two \ce{H2} molecules are on top or bottom of the ligand-\ce{TmCl_x} complex and weakly polarized toward the \ce{Cl^-} ions (Figure \ref{fig:3Dmolecules}d). With the addition of the third and fourth \ce{H2} molecules, the location of \ce{H2} molecules rearrange mostly into typical configurations (Figure \ref{fig:3Dmolecules}e-i). Transition metals with higher oxidation states: Cr(III), Sc(III), V(V), and Ti(IV)  form square pyramidal, trigonal bipyramidal, octahedral, or  pentagonal bipyramidal geometries, respectively. The first \ce{H2} molecule has a minimum energy at the nearest possible location to the Tm and binding sites (N or O atoms). Some Tm have different geometries depending on the bond length and ligand such as in the ligand-\ce{VCl_5} complexes. V(V) has pentagonal bipyramidal geometry in \textbf{PIA}, \textbf{BPY}, and \textbf{PHEN} while octahedral geometry in \textbf{BBH} and \textbf{PIP}. In general, the distance between \ce{H2} to the Tm in the pentagonal bipiramidal (Pbi) and octahedral (Oct) geometries are about 1 \textrm{\AA} longer than other geometries because of  the smaller amount of available space for the \ce{H2} to interact directly with the Tm, and consequently have lower binding enthalpy. In the trigonal bypiramidal Sc(III) and square pyramidal Cr(III) configurations, the binding enthalpies are comparable to  the square planar and tetrahedral geometries, respectively. In the case of Sc(III) or Cr(III), the first \ce{H2} have strong binding enthalpy because it can get close ($\sim$ 2.8 \AA) to the Tm centers and thus more interactions occur.

The $3^{rd}$ studied parameter is the effect of electrostatics and dispersion interactions between the \ce{H2} molecules and the linker-\ce{TmCl_x} complexes. The leading permanent multipole moment of the \ce{H2} is a weak quadropole moment but it could also have weak induced dipole moment.\cite{Lochan_PCCP_2006,stern_understanding_2012,Pascal_JPCL_2011}  A fragment analysis proposed by Ziegler and Berends has been performed to decompose the binding enthalpy. \cite{bickelhaupt_kohn-sham_2000,ziegler_calculation_1977} To illustrate this we showed two cases: 
the \ce{H2} with a quadrupole moment is attracted to Cu(II) more than  to V(V) by about 2.8 kJ/mol. In all of these compounds, the effect of electrostatics is 5\% smaller than the dispersion. The dispersion energy, electrostatics, Pauli repulsions and orbital interactions in the binding enthalpy between the Tm-ligand complexes and \ce{H2} for \textbf{BPY} are shown in Figure \ref{fig:distanceTm_H2}b. This analysis  shows that the dispersion energy and electrostatics are the dominant factors for the magnitude of the binding enthalpy. The other ligands  follow a similar trend as \textbf{BPY} (Supporting Information section 2). 

\begin{figure}[htbp!]
	\includegraphics[width=.99\linewidth]{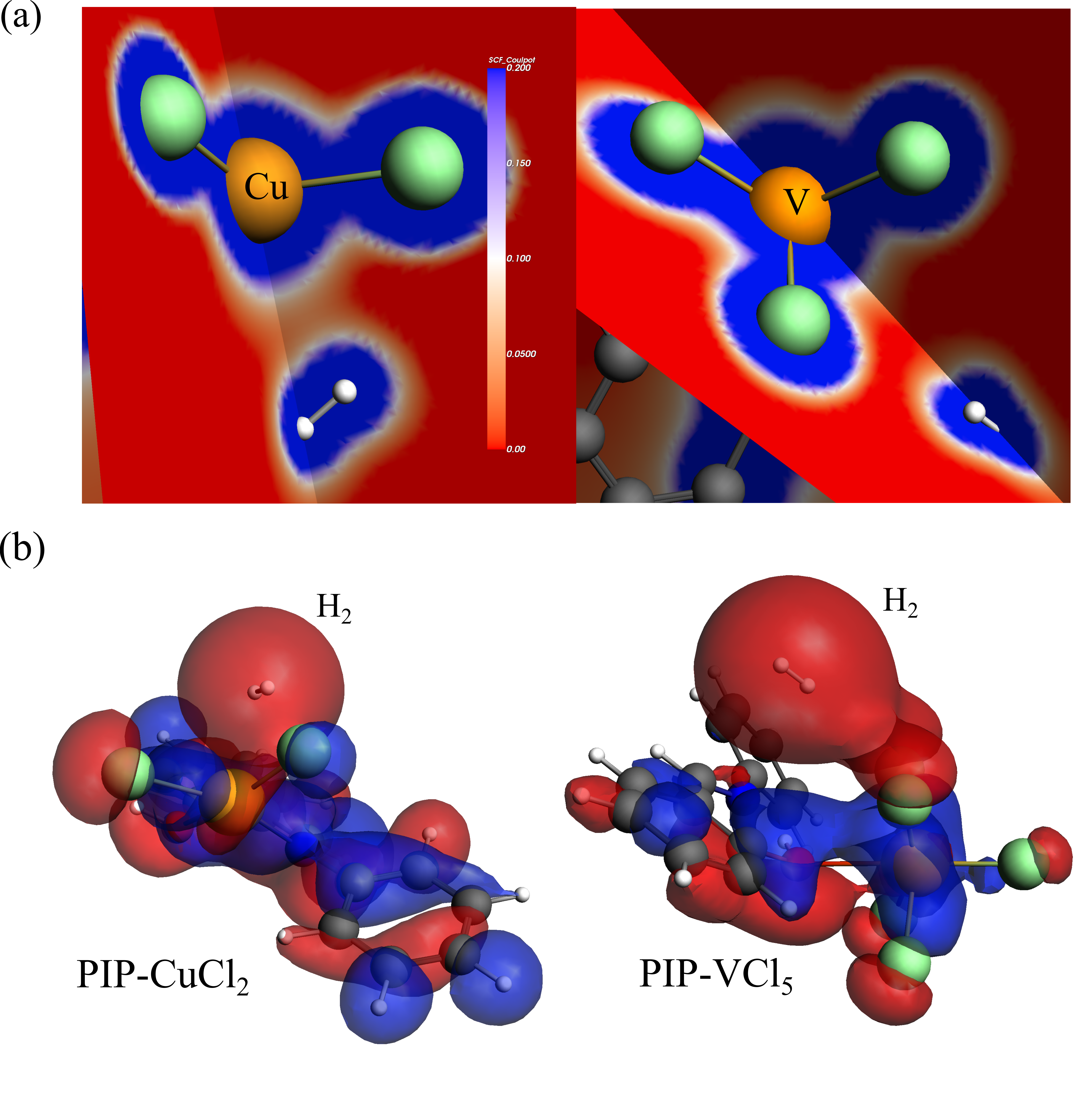}
	\caption{\label{fig:ElectronClouds} HOMO-LUMO orbitals projected in (a) planes and (b) space for \textbf{PIP}-\ce{CuCl2}+\ce{H2} and \textbf{PIP}-\ce{VCl5}+\ce{H2} complexes. In the former the \ce{H2} interacts mostly with Cu(II) while in the later, the \ce{H2} interacts more with \ce{Cl^-} than with the V(V) atom.}
\end{figure}

The $4^{th}$ studied parameter is the possible orbital interactions between the s-orbitals of \ce{H2} and the d-orbitals of the Tm. This can be estimated by the occupied molecular orbitals (MO) of the complexes in Figure \ref{fig:ElectronClouds}. In Figure \ref{fig:ElectronClouds}b the occupied MO has some overlap between the s-orbital from \ce{H2} and d-orbital of Cu(II). A similar phenomenon is observed in the \textbf{PIP}-\ce{VCl5} complex. The orbital interaction depends on the overlap between the interacting orbitals and decays exponentially with the distance between \ce{H2} and ligand-\ce{TmCl_x} complex.\cite{Lochan_PCCP_2006} 

\begin{figure}[htbp!]
	\includegraphics[width=.99\linewidth]{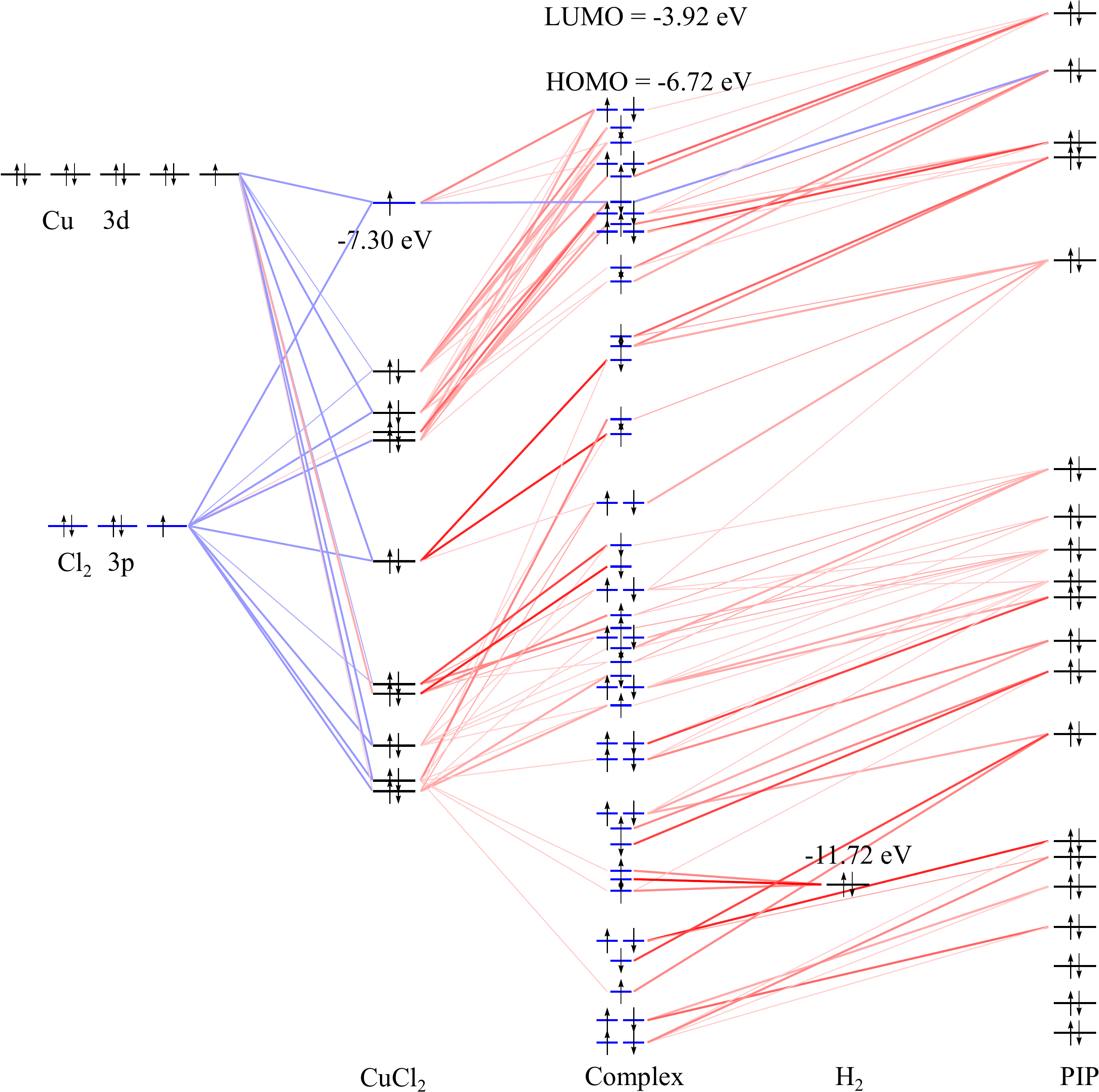}
	\caption{\label{fig:FragmentLevel} The molecular orbitals (MO) of the complex with  linker-\ce{CuCl2} and one \ce{H2}. The components of the complex are decomposed as \ce{Cu}, \ce{Cl2}, \ce{CuCl2}, and linker \textbf{PIP}. The MO at the level of $\approx$-11.72 eV are composed mainly from the s-orbitals of \ce{H2} and some contribution from the d-orbital of transition metal Cu(II).} 
\end{figure}

Fragment analysis has been performed to differentiate the molecular orbitals (MO) from each components of the complex (\ce{H2}, linker, and \ce{TmCl_x}). The fragment analysis also computes the contributing components of total energy. The s-orbitals of \ce{H2} molecules are overlapped with the d-orbitals of Tm atoms in the linker-\ce{TmCl_x} complexes; \textit{e.g.} about 70\% contribution of molecular orbitals from \ce{H2} s-orbitals are overlapped with the d-orbitals of the Cu atom in the \textbf{PIP}-\ce{CuCl2} complex (Figure \ref{fig:ElectronClouds}). Typically, the energy level of the occupied MO (HOMO) of \ce{H2} molecule alone (-11.72 eV) is far below the HOMO (-6.72 eV) of the whole system; \textit{e.g.} \ce{CuCl2}-\textbf{PIP} + \ce{H2} (Figure  \ref{fig:FragmentLevel}). This is consistent with the \ce{H2} as a poor charge donor because of the deep energy level of its $\sigma$ bonding energy level $-11.72 \,\mathrm{eV}$ and also a poor charge acceptor because of the high level of its $\sigma^*$ anti-bonding energy level (beyond the scale of Figure \ref{fig:FragmentLevel}).

The profiles of HOMO and LUMO gap of all the complexes is shown in Fig \ref{fig:distanceTm_H2}c. The energy gap of the pure ligand is higher than any chelated complex.  The ligand-\ce{VCl5} complexes have the lowest gap at $\sim$1.2 eV. Most of the complexes have an energy gap in the visible light spectrum (1.6 - 3.2 eV), therefore these complexes are promising candidates as dye sensitizers.  

In summary, the nature of intermolecular interactions between \ce{H2}  and  linker-\ce{TmCl_x} complexes is presented. The favorable sites for the \ce{H2} molecules interactions depend mainly on the type of the Tm, spin, and available space. An important conclusion of this work concerns the design of chelated linkers for crystalline porous materials such as COFs and MOFs. The chelation of Tm inside porous frameworks through their linkers can enhance the effective \ce{H2} storage as the interaction between the Tm atoms and \ce{H2} can be tuned to get higher binding enthalpy  $\Delta \textit{H}_{\textrm{bind}}^{\circ}$. Ultimately, the ability of the chelated linkers to bind \ce{H2} molecules is highly dependent on the Tm coordination sphere, with the most important interaction given by the dispersion and electrostatics. 

\textit{Supplementary Information}. Computational details and chelated complexes with the \ce{H2} optimized linker structures and their descriptions are provided. Optimized Geometries of the chelated complexes with \ce{H2} are available free of charge on the ACS Publications website.    

\begin{acknowledgement}
	
J.L.M-C. was supported by Florida State University (FSU). The authors gratefully acknowledge the support from the Energy and Materials Initiative and the High Performance Material Institute (HPMI) facilities at FSU. S.P. thanks the Science and Engineering Research Board-Department of Science and Technology (SERB-DST), Government of India for providing the Ramanujan Faculty Fellowship under the Grant No. SB/S2/RJN-067/2017. A portion of this work was performed at the National High Magnetic Field Laboratory, which is supported by National Science Foundation Cooperative Agreement No. DMR-1644779 and the State of Florida. The authors thank the High Performance Computer cluster at the Research Computing Center (RCC) at FSU for providing computational resources and support.   \\
	
\end{acknowledgement}

\bibliography{bibliography_H2linkers}

\providecommand{\latin}[1]{#1}
\makeatletter
\providecommand{\doi}
  {\begingroup\let\do\@makeother\dospecials
  \catcode`\{=1 \catcode`\}=2 \doi@aux}
\providecommand{\doi@aux}[1]{\endgroup\texttt{#1}}
\makeatother
\providecommand*\mcitethebibliography{\thebibliography}
\csname @ifundefined\endcsname{endmcitethebibliography}
  {\let\endmcitethebibliography\endthebibliography}{}
\begin{mcitethebibliography}{28}
\providecommand*\natexlab[1]{#1}
\providecommand*\mciteSetBstSublistMode[1]{}
\providecommand*\mciteSetBstMaxWidthForm[2]{}
\providecommand*\mciteBstWouldAddEndPuncttrue
  {\def\EndOfBibitem{\unskip.}}
\providecommand*\mciteBstWouldAddEndPunctfalse
  {\let\EndOfBibitem\relax}
\providecommand*\mciteSetBstMidEndSepPunct[3]{}
\providecommand*\mciteSetBstSublistLabelBeginEnd[3]{}
\providecommand*\EndOfBibitem{}
\mciteSetBstSublistMode{f}
\mciteSetBstMaxWidthForm{subitem}{(\alph{mcitesubitemcount})}
\mciteSetBstSublistLabelBeginEnd
  {\mcitemaxwidthsubitemform\space}
  {\relax}
  {\relax}

\bibitem[Lucht and Mendoza-Cortes(2015)Lucht, and Mendoza-Cortes]{Lucht2015}
Lucht,~K.~P.; Mendoza-Cortes,~J.~L. {Birnessite: A Layered Manganese Oxide To
  Capture Sunlight for Water-Splitting Catalysis}. \emph{J. Phys. Chem. C}
  \textbf{2015}, \emph{119}\relax
\mciteBstWouldAddEndPuncttrue
\mciteSetBstMidEndSepPunct{\mcitedefaultmidpunct}
{\mcitedefaultendpunct}{\mcitedefaultseppunct}\relax
\EndOfBibitem
\bibitem[Pramudya and Mendoza-Cortes(2016)Pramudya, and
  Mendoza-Cortes]{mendoza_design_2016}
Pramudya,~Y.; Mendoza-Cortes,~J.~L. \emph{J. Am. Chem. Soc.} \textbf{2016},
  \emph{138}, 15204--15213\relax
\mciteBstWouldAddEndPuncttrue
\mciteSetBstMidEndSepPunct{\mcitedefaultmidpunct}
{\mcitedefaultendpunct}{\mcitedefaultseppunct}\relax
\EndOfBibitem
\bibitem[Yildirim and Hartman(2005)Yildirim, and Hartman]{yildirim_direct_2005}
Yildirim,~T.; Hartman,~M.~R. Direct {Observation} of {Hydrogen} {Adsorption}
  {Sites} and {Nanocage} {Formation} in {Metal}-{Organic} {Frameworks}.
  \emph{Phys. Rev. Lett.} \textbf{2005}, \emph{95}, 215504\relax
\mciteBstWouldAddEndPuncttrue
\mciteSetBstMidEndSepPunct{\mcitedefaultmidpunct}
{\mcitedefaultendpunct}{\mcitedefaultseppunct}\relax
\EndOfBibitem
\bibitem[Arter \latin{et~al.}(2016)Arter, Zuluaga, Harrison, Welchman, and
  Thonhauser]{arter_fivefold_2016}
Arter,~C.~A.; Zuluaga,~S.; Harrison,~D.; Welchman,~E.; Thonhauser,~T. Fivefold
  increase of hydrogen uptake in {MOF}74 through linker decorations.
  \emph{Phys. Rev. B} \textbf{2016}, \emph{94}, 144105\relax
\mciteBstWouldAddEndPuncttrue
\mciteSetBstMidEndSepPunct{\mcitedefaultmidpunct}
{\mcitedefaultendpunct}{\mcitedefaultseppunct}\relax
\EndOfBibitem
\bibitem[Aduenko \latin{et~al.}(2018)Aduenko, Murray, and
  Mendoza-Cortes]{Aduenko2018}
Aduenko,~A.~A.; Murray,~A.; Mendoza-Cortes,~J.~L. General Theory of Absorption
  in Porous Materials: Restricted Multilayer Theory. \emph{ACS Appl. Mater.
  Interfaces} \textbf{2018}, \emph{10}, 13244--13251\relax
\mciteBstWouldAddEndPuncttrue
\mciteSetBstMidEndSepPunct{\mcitedefaultmidpunct}
{\mcitedefaultendpunct}{\mcitedefaultseppunct}\relax
\EndOfBibitem
\bibitem[Han and Goddard(2007)Han, and Goddard]{Han_JACS_2007}
Han,~S.~S.; Goddard,~W.~A. Lithium-doped metal-organic frameworks for
  reversible H-2 storage at ambient temperature. \emph{J. Am. Chem. Soc.}
  \textbf{2007}, \emph{129}, 8422--8423\relax
\mciteBstWouldAddEndPuncttrue
\mciteSetBstMidEndSepPunct{\mcitedefaultmidpunct}
{\mcitedefaultendpunct}{\mcitedefaultseppunct}\relax
\EndOfBibitem
\bibitem[Bhatia and Myers(2006)Bhatia, and Myers]{Bhatia_Langmuir_2006}
Bhatia,~S.~K.; Myers,~A.~L. Optimum conditions for adsorptive storage.
  \emph{Langmuir} \textbf{2006}, \emph{22}, 1688--1700\relax
\mciteBstWouldAddEndPuncttrue
\mciteSetBstMidEndSepPunct{\mcitedefaultmidpunct}
{\mcitedefaultendpunct}{\mcitedefaultseppunct}\relax
\EndOfBibitem
\bibitem[Lochan and Head-Gordon(2006)Lochan, and Head-Gordon]{Lochan_PCCP_2006}
Lochan,~R.~C.; Head-Gordon,~M. Computational studies of molecular hydrogen
  binding affinities: The role of dispersion forces, electrostatics, and
  orbital interactions. \emph{Phys. Chem. Chem. Phys.} \textbf{2006}, \emph{8},
  1357--1370\relax
\mciteBstWouldAddEndPuncttrue
\mciteSetBstMidEndSepPunct{\mcitedefaultmidpunct}
{\mcitedefaultendpunct}{\mcitedefaultseppunct}\relax
\EndOfBibitem
\bibitem[Pascal \latin{et~al.}(2011)Pascal, Boxe, and
  Goddard]{Pascal_JPCL_2011}
Pascal,~T.~A.; Boxe,~C.; Goddard,~W.~A. An Inexpensive, Widely Available
  Material for 4 wt \% Reversible Hydrogen Storage Near Room Temperature.
  \emph{J. Phys. Chem. Lett.} \textbf{2011}, \emph{2}, 1417--1420\relax
\mciteBstWouldAddEndPuncttrue
\mciteSetBstMidEndSepPunct{\mcitedefaultmidpunct}
{\mcitedefaultendpunct}{\mcitedefaultseppunct}\relax
\EndOfBibitem
\bibitem[Kubas(2007)]{Kubas_ChemRev_2007}
Kubas,~G.~J. Fundamentals of H(2) binding and reactivity on transition metals
  underlying hydrogenase function and H(2) production and storage. \emph{Chem.
  Rev.} \textbf{2007}, \emph{107}, 4152--4205\relax
\mciteBstWouldAddEndPuncttrue
\mciteSetBstMidEndSepPunct{\mcitedefaultmidpunct}
{\mcitedefaultendpunct}{\mcitedefaultseppunct}\relax
\EndOfBibitem
\bibitem[Becke(1988)]{Becke1988}
Becke,~A.~D. {Density-Functional Exchange-energy Approximation with Correct
  Asymptotic Behavior}. \emph{Phys. Rev. A} \textbf{1988}, \emph{38},
  3098--3100\relax
\mciteBstWouldAddEndPuncttrue
\mciteSetBstMidEndSepPunct{\mcitedefaultmidpunct}
{\mcitedefaultendpunct}{\mcitedefaultseppunct}\relax
\EndOfBibitem
\bibitem[Lee \latin{et~al.}(1988)Lee, Yang, and Parr]{Lee1988}
Lee,~C.; Yang,~W.; Parr,~R.~G. {Development of the Colle-Salvetti
  Correlation-Energy Formula into a Functional of the Electron Density}.
  \emph{Phys. Rev. B} \textbf{1988}, \emph{37}, 785--789\relax
\mciteBstWouldAddEndPuncttrue
\mciteSetBstMidEndSepPunct{\mcitedefaultmidpunct}
{\mcitedefaultendpunct}{\mcitedefaultseppunct}\relax
\EndOfBibitem
\bibitem[Grimme(2006)]{Grimme2006}
Grimme,~S. {Semiempirical GGA-type Density Functional Constructed with a
  Long-range Dispersion Correction}. \emph{J. Comput. Chem.} \textbf{2006},
  \emph{27}, 1787--1799\relax
\mciteBstWouldAddEndPuncttrue
\mciteSetBstMidEndSepPunct{\mcitedefaultmidpunct}
{\mcitedefaultendpunct}{\mcitedefaultseppunct}\relax
\EndOfBibitem
\bibitem[Grimme \latin{et~al.}(2010)Grimme, Antony, Ehrlich, and
  Krieg]{Grimme2010}
Grimme,~S.; Antony,~J.; Ehrlich,~S.; Krieg,~H. \emph{The Journal of Chemical
  Physics} \textbf{2010}, \emph{132}, 154104\relax
\mciteBstWouldAddEndPuncttrue
\mciteSetBstMidEndSepPunct{\mcitedefaultmidpunct}
{\mcitedefaultendpunct}{\mcitedefaultseppunct}\relax
\EndOfBibitem
\bibitem[Pakhira and Mendoza-Cortes(2018)Pakhira, and
  Mendoza-Cortes]{Jose2018jpcc}
Pakhira,~S.; Mendoza-Cortes,~J.~L. Tuning Dirac Cone of Two Dimensional Bilayer
  Graphene and Graphite by Intercalating First Row Transition Metals using
  First Principles. \emph{J. Phys. Chem. C} \textbf{2018}, \emph{122},
  4768--4782\relax
\mciteBstWouldAddEndPuncttrue
\mciteSetBstMidEndSepPunct{\mcitedefaultmidpunct}
{\mcitedefaultendpunct}{\mcitedefaultseppunct}\relax
\EndOfBibitem
\bibitem[Pakhira \latin{et~al.}(2017)Pakhira, Lucht, and
  Mendoza-Cortes]{Pakhira2017c}
Pakhira,~S.; Lucht,~K.~P.; Mendoza-Cortes,~J.~L. {Iron Intercalated
  Covalent-Organic Frameworks: A Promising Approach for Semiconductors}.
  \emph{J. Phys. Chem. C} \textbf{2017}, \emph{121}, 21160--21170\relax
\mciteBstWouldAddEndPuncttrue
\mciteSetBstMidEndSepPunct{\mcitedefaultmidpunct}
{\mcitedefaultendpunct}{\mcitedefaultseppunct}\relax
\EndOfBibitem
\bibitem[Pakhira \latin{et~al.}(2015)Pakhira, Takayanagi, and
  Nagaoka]{Masataka2015a}
Pakhira,~S.; Takayanagi,~M.; Nagaoka,~M. {Diverse Rotational Flexibility of
  Substituted Dicarboxylate Ligands in Functional Porous Coordination
  Polymers}. \emph{J. Phys. Chem. C} \textbf{2015}, \emph{119},
  28789--28799\relax
\mciteBstWouldAddEndPuncttrue
\mciteSetBstMidEndSepPunct{\mcitedefaultmidpunct}
{\mcitedefaultendpunct}{\mcitedefaultseppunct}\relax
\EndOfBibitem
\bibitem[Hui \latin{et~al.}(2018)Hui, Pakhira, Bhargava, Barton, Zhou,
  Chinderle, Mendoza-Cortes, and Rodriguez-Lopez]{Jingshu2018}
Hui,~J.; Pakhira,~S.; Bhargava,~R.; Barton,~Z.~J.; Zhou,~X.; Chinderle,~A.~J.;
  Mendoza-Cortes,~J.~L.; Rodriguez-Lopez,~J. Modulating Electrocatalysis on
  Graphene Heterostructures: Physically Impermeable Yet Electronically
  Transparent Electrodes. \emph{ACS Nano} \textbf{2018}, \emph{12},
  2980--2990\relax
\mciteBstWouldAddEndPuncttrue
\mciteSetBstMidEndSepPunct{\mcitedefaultmidpunct}
{\mcitedefaultendpunct}{\mcitedefaultseppunct}\relax
\EndOfBibitem
\bibitem[Niu \latin{et~al.}(2018)Niu, Pakhira, Marcus, Li, Mendoza-Cortes, and
  Yang]{Niu2018x}
Niu,~W.; Pakhira,~S.; Marcus,~K.; Li,~Z.; Mendoza-Cortes,~J.~L.; Yang,~Y.
  Apically Dominant Mechanism for Improving Catalytic Activities of N-Doped
  Carbon Nanotube Arrays in Rechargeable Zinc--Air Battery. \emph{Adv. Ene.
  Mat.} \textbf{2018}, 1800480\relax
\mciteBstWouldAddEndPuncttrue
\mciteSetBstMidEndSepPunct{\mcitedefaultmidpunct}
{\mcitedefaultendpunct}{\mcitedefaultseppunct}\relax
\EndOfBibitem
\bibitem[Pakhira \latin{et~al.}(2018)Pakhira, Lucht, and
  Mendoza-Cortes]{Jose2018jcp}
Pakhira,~S.; Lucht,~K.~P.; Mendoza-Cortes,~J.~L. Dirac Cone in two dimensional
  bilayer graphene by intercalation with V, Nb, and Ta transition metals.
  \emph{J. Chem. Phys.} \textbf{2018}, \emph{148}, 064707\relax
\mciteBstWouldAddEndPuncttrue
\mciteSetBstMidEndSepPunct{\mcitedefaultmidpunct}
{\mcitedefaultendpunct}{\mcitedefaultseppunct}\relax
\EndOfBibitem
\bibitem[Ghosh \latin{et~al.}(2008)Ghosh, Biswas, Florke, and
  Nag]{Ghosh_InorgChem_2008}
Ghosh,~M.; Biswas,~P.; Florke,~U.; Nag,~K. Halogen exchange and scrambling
  between C-X and M-X ' bonds in copper, nickel, and cobalt complexes of 6,6
  '-bis(bromo/chloromethyl)-2,2 '-bipyridine. Structural, electrochemical, and
  photochemical studies. \emph{Inorg. Chem.} \textbf{2008}, \emph{47},
  281--296\relax
\mciteBstWouldAddEndPuncttrue
\mciteSetBstMidEndSepPunct{\mcitedefaultmidpunct}
{\mcitedefaultendpunct}{\mcitedefaultseppunct}\relax
\EndOfBibitem
\bibitem[Wu \latin{et~al.}(1999)Wu, Janiak, Rheinwald, and
  Land]{Wu_JChemSoc_1999}
Wu,~H.~P.; Janiak,~C.; Rheinwald,~G.; Land,~H. 5,5-dicyano-2,2 '-bipyridine
  silver complexes: discrete units or co-ordination polymers through a
  chelating and/or bridging metal-ligand interaction. \emph{J. Chem. Soc.
  Dalton Trans.} \textbf{1999}, 183--190\relax
\mciteBstWouldAddEndPuncttrue
\mciteSetBstMidEndSepPunct{\mcitedefaultmidpunct}
{\mcitedefaultendpunct}{\mcitedefaultseppunct}\relax
\EndOfBibitem
\bibitem[Thonhauser \latin{et~al.}(2015)Thonhauser, Zuluaga, Arter, Berland,
  Schröder, and Hyldgaard]{thonhauser_spin_2015}
Thonhauser,~T.; Zuluaga,~S.; Arter,~C.; Berland,~K.; Schröder,~E.;
  Hyldgaard,~P. Spin {Signature} of {Nonlocal} {Correlation} {Binding} in
  {Metal}-{Organic} {Frameworks}. \emph{Phys. Rev. Lett.} \textbf{2015},
  \emph{115}, 136402\relax
\mciteBstWouldAddEndPuncttrue
\mciteSetBstMidEndSepPunct{\mcitedefaultmidpunct}
{\mcitedefaultendpunct}{\mcitedefaultseppunct}\relax
\EndOfBibitem
\bibitem[Sun \latin{et~al.}(2007)Sun, Kim, and Zhang]{sun_effect_2007}
Sun,~Y.~Y.; Kim,~Y.-H.; Zhang,~S.~B. Effect of {Spin} {State} on the
  {Dihydrogen} {Binding} {Strength} to {Transition} {Metal} {Centers} in
  {Metal}-{Organic} {Frameworks}. \emph{J. Am. Chem. Soc.} \textbf{2007},
  \emph{129}, 12606--12607\relax
\mciteBstWouldAddEndPuncttrue
\mciteSetBstMidEndSepPunct{\mcitedefaultmidpunct}
{\mcitedefaultendpunct}{\mcitedefaultseppunct}\relax
\EndOfBibitem
\bibitem[Stern \latin{et~al.}(2012)Stern, Belof, Eddaoudi, and
  Space]{stern_understanding_2012}
Stern,~A.~C.; Belof,~J.~L.; Eddaoudi,~M.; Space,~B. Understanding hydrogen
  sorption in a polar metal-organic framework with constricted channels.
  \emph{J. Chem. Phys.} \textbf{2012}, \emph{136}, 034705\relax
\mciteBstWouldAddEndPuncttrue
\mciteSetBstMidEndSepPunct{\mcitedefaultmidpunct}
{\mcitedefaultendpunct}{\mcitedefaultseppunct}\relax
\EndOfBibitem
\bibitem[Bickelhaupt and Baerends(2000)Bickelhaupt, and
  Baerends]{bickelhaupt_kohn-sham_2000}
Bickelhaupt,~F.~M.; Baerends,~E.~J. \emph{Reviews in {Computational}
  {Chemistry}}; John Wiley \& Sons, Inc., Hoboken, 2000; p~1\relax
\mciteBstWouldAddEndPuncttrue
\mciteSetBstMidEndSepPunct{\mcitedefaultmidpunct}
{\mcitedefaultendpunct}{\mcitedefaultseppunct}\relax
\EndOfBibitem
\bibitem[Ziegler and Rauk(1977)Ziegler, and Rauk]{ziegler_calculation_1977}
Ziegler,~T.; Rauk,~A. On the calculation of bonding energies by the
  Hartree-Fock Slater method. \emph{Theoret. Chim. Acta} \textbf{1977},
  \emph{46}, 1--10\relax
\mciteBstWouldAddEndPuncttrue
\mciteSetBstMidEndSepPunct{\mcitedefaultmidpunct}
{\mcitedefaultendpunct}{\mcitedefaultseppunct}\relax
\EndOfBibitem
\end{mcitethebibliography}

\clearpage
\newpage
TABLE OF CONTENTS GRAPHIC

\begin{figure}
\centering
\includegraphics[width=0.99\linewidth]{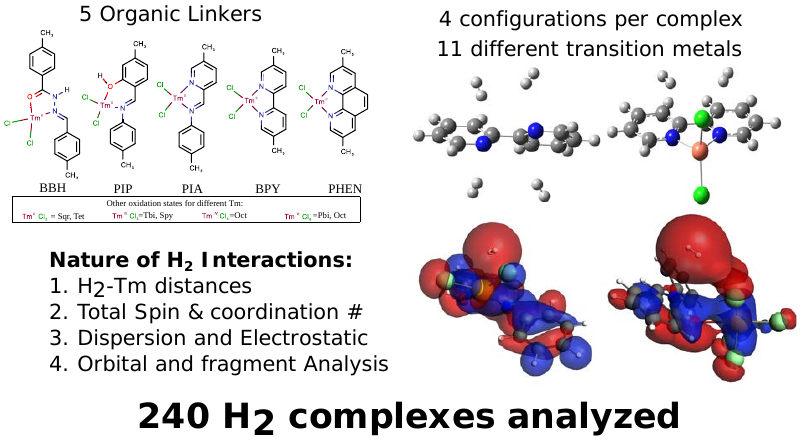}
\end{figure}
 
\end{document}